\documentclass{PoS}

\title{Nonlinear interactions for massive spin-2 fields}

\ShortTitle{Nonlinear interactions for massive spin-2 fields}

\author{\speaker{Angnis Schmidt-May}
\\
        ETH Z\"urich\\ Institute for Theoretical Physics\\ Wolfgang-Pauli-Str. 27\\ CH - 8093 Z\"urich\\
        E-mail: \email{angniss@itp.phys.ethz.ch}}

\abstract{We give a basic introduction to ghost-free nonlinear theories involving massive spin-2 fields, focussing on bimetric theory. After motivating the construction of such models from field theoretical considerations, we review the linear theories for massive and massless spin-2 fluctuations propagating on maximally symmetric backgrounds. The structure of general nonlinear spin-2 interactions is explained before we specialise to the ghost-free case. We review the maximally symmetric solutions of bimetric theory, its mass spectrum and the parameter limit which brings the theory close to general relativity. Finally we discuss applications of bimetric theory to cosmology with particular emphasis on the role of the general relativity limit.}

\FullConference{Proceedings of the Corfu Summer Institute 2015 "School and Workshops on Elementary Particle Physics and Gravity"\\
		1-27 September 2015\\
		Corfu, Greece}

\newcommand{\gmn}{g_{\mu\nu}}
\newcommand{\fmn}{f_{\mu\nu}}

\newcommand{\nn}{\nonumber}
\newcommand{\beqn}{\begin{eqnarray}}
\newcommand{\eeqn}{\end{eqnarray}}
\newcommand{\dd}{\mathrm{d}}
\newcommand{\Tr}{\mathrm{Tr}}

\begin{document}
%%%%%%%%%%%%%%%%%%%%%%%%%%%%%%%%%%%%%%%%%%%%%%%%%%%%%%%%%%%%%%%%
%%%%%%%%%%%%%%%%%%%%%%%%%%%%%%%%%%%%%%%%%%%%%%%%%%%%%%%%%%%%%%%%
%%%%%%%%%%%%%%%%%%%%%%%%%%%%%%%%%%%%%%%%%%%%%%%%%%%%%%%%%%%%%%%%

%%%%%%%%%%%%%%%%%%%%%%%%%%%%%%%%%%%%%%%%%%%%%%%%%%%%%%%%%%%%%%%%
\section{Motivation}
%%%%%%%%%%%%%%%%%%%%%%%%%%%%%%%%%%%%%%%%%%%%%%%%%%%%%%%%%%%%%%%%

At present, the $\Lambda$-cold-dark-matter ($\Lambda$CDM) model is our best framework for describing physics at cosmological scales~\cite{Adam:2015rua}. It is based on general relativity (GR) as the theory for gravity, including a cosmological constant that gives rise to the observed acceleration of the cosmic expansion. In addition to the particles of the Standard Model, one assumes the presence of a dark matter component in the matter sector whose precise properties (spin, mass, gauge charges) are unknown. 
Cosmologists strive for a better understanding of the two most challenging problems faced by the $\Lambda$CDM model:
disclosing the obscure nature of the dark matter particle(s) and explaining the observed value for the cosmological constant which is unnaturally small compared to expectations from quantum field theory.

Most theoretical models aiming to explain one or both of the unresolved puzzles involve new degrees of freedom, either in the gravitational or in the matter sector. For instance, an extra scalar on the gravity side can give rise to cosmic acceleration in the absence of a cosmological constant term or, in supersymmetric extensions of the Standard Model, a super-partner of a known particle can act as dark matter candidate. The additional degrees of freedom invoked in such extended models are usually new versions of fields already present in the Standard Model, i.e.~they are massive or massless fields of spin 0, $\frac{1}{2}$ or 1 with some assigned gauge charges. The theoretical framework for such fields is well-understood (see Table 1) and there is a large number of possibilities to extend the $\Lambda$CDM model by such degrees of freedom. Most of the extended models are subject to strong phenomenological constraints. In particular, GR is extremely well-tested in the solar system and confirmed to give a precise description of gravity down to very small distance scales. Hence, any modification of the gravitational sector needs to be such that the new theory resembles GR over a large range of energies. This means that either the additional degrees of freedom can at most couple very weakly to standard matter fields or their influence on the latter needs to be diminished by a screening mechanism. Consequently, it is not clear if any new effects can solve the problems of standard cosmology without violating other tests of GR. 

\begin{table}[hdtp]
\begin{center}
\begin{tabular}{c|c|c}
spin 0& scalar field $\phi$ & $\mathcal{L}_\phi=-\partial_\mu\phi\partial^\mu\phi-m^2\phi^2$
 %\\ &&
 \\
\hline && \\
spin $\frac{1}{2}$&  spinor field $\psi^\alpha$ & $\mathcal{L}_\psi=-\bar{\psi}\gamma^\mu\partial_\mu\psi-m\bar{\psi}\psi$ 
%\\ &&
\\
\hline &&\\
spin 1& vector field $A_\mu$& $\mathcal{L}_A=-\frac{1}{4}F^{\mu\nu}F_{\mu\nu}-\frac{m^2}{2}A^\mu A_\mu$ 
%\\ &&
%\\ \hline 
\end{tabular}
\end{center}
\caption{Free Lagrangians for fields of different spin. The massless theories correspond to setting $m=0$.}
\end{table}

An alternative approach is to put the phenomenological issues aside for the time being and inspect the ingredients of the standard cosmological model and its extensions from a more theoretical point of view. One immediately notices that, from the field theoretical perspective, the set of models that are usually considered appears to be incomplete. While lower spin fields can be either massless or massive, the spin-2 graviton of GR is always massless. The obvious question that arises is whether there exist models including massive spin-2 degrees of freedom that could be realised in nature. A subtlety, which is not encountered in the lower-spin cases, already occurs in the linear theory for massive spin-2 fields. On top of that, since GR is highly nonlinear, the mass term needs to be completed into a nonlinear interaction potential.
Finding such self-interactions for tensor fields like the metric $\gmn$ turns out to be surprisingly difficult.

The study of massive spin-2 fields was initiated already in 1939 by Fierz and Pauli who wrote down the linear equations in flat space~\cite{Fierz:1939ix}. The search for a nonlinear completion of the spin-2 mass term was interrupted in 1972 when Boulware and Deser demonstrated the presence of an inevitable ghost instability in generic types of self-interactions for the massive spin-2 field~\cite{Boulware:1973my}. This result was received as a reliable no-go theorem and for a long time there was little development in the field.
It was only in 2011 that de Rham, Gabadadze and Tolley discovered a loophole in the assumptions entering the proof of Boulware and Deser.  Based on previous analyses carried out by other groups in~\cite{ArkaniHamed:2002sp, Creminelli:2005qk}, they presented a candidate for a consistent nonlinear spin-2 mass term~\cite{deRham:2010ik, deRham:2010kj}. Shortly after, Hassan and Rosen confirmed that the proposed theory of massive gravity indeed avoids the Boulware-Deser ghost instability~\cite{Hassan:2011hr, Hassan:2011tf} and then generalised it to a fully dynamical and ghost-free bimetric theory~\cite{Hassan:2011zd}. Since then, a lot of progress has been made in the field of nonlinear spin-2 theories and their application to cosmology. Here we can only give a brief overview of the subject and for more details we refer to several recent reviews focussing on massive gravity~\cite{Hinterbichler:2011tt, deRham:2014zqa} and bimetric theory~\cite{Schmidt-May:2015vnx}.

%%%%%%%%%%%%%%%%%%%%%%%%%%%%%%%%%%%%%%%%%%%%%%%%%%%%%%%%%%%%%%%%
\section{Massless and massive spin-2 fields}
%%%%%%%%%%%%%%%%%%%%%%%%%%%%%%%%%%%%%%%%%%%%%%%%%%%%%%%%%%%%%%%%

 % % % % % % % % % % % % % % % % % % % % % % % % % % % % % %
\subsection{Massless gravity}
 % % % % % % % % % % % % % % % % % % % % % % % % % % % % % %

General relativity is a nonlinear field theory for a symmetric rank-2 tensor field $\gmn$ whose dynamics are dictated by the Einstein-Hilbert action,
\beqn
S_\mathrm{EH}= m_\mathrm{P}^2\int\dd^4x~\sqrt{g}~(R-2\Lambda)\,.
\eeqn
Here $m_\mathrm{P}$ is the Planck mass, $g=|\det \gmn|$ and $R$ is the Ricci curvature scalar for the metric tensor $\gmn$. We have also included a cosmological constant term $\Lambda$ into the action. The above expression is the most general action, involving only $\gmn$ and at most two derivatives, which is invariant under general coordinate transformations (GCT) with gauge parameters $\xi_\mu$. Infinitesimally, these transformations act on the metric as $\gmn\mapsto\gmn +\nabla_\mu\xi_\nu+\nabla_\nu\xi_\mu$, where $\nabla$ is the covariant derivative compatible with $\gmn$.

The gravitational interactions of matter fields (collectively denoted by $\Phi_i$) arise from a covariantised matter action of the form,
\beqn\label{matter}
S_\mathrm{m}=\int\dd^4x~\mathcal{L}_\mathrm{m}(g, \Phi_i)\,.
\eeqn
For instance, for a free massless scalar $\phi$ we would have $\mathcal{L}_\mathrm{m}(g, \phi)=-\sqrt{g}~g^{\mu\nu}\partial_\mu\phi \partial_\nu \phi$.
Varying the Einstein-Hilbert action including the matter coupling with respect to $\gmn$ results in Einstein's equations,
\beqn\label{EE}
R_{\mu\nu}-\frac{1}{2}\gmn R+\Lambda\gmn=\frac{1}{m_\mathrm{P}^2}T_{\mu\nu}\,,
\eeqn
with stress-energy tensor $T_{\mu\nu}=-\frac{1}{\sqrt{g}}\frac{\delta \mathcal{L}_\mathrm{m}}{\delta g^{\mu\nu}}$. In the following we focus on Einstein's equation in vacuum, i.e. with~$T_{\mu\nu}=0$, in which case their solutions $\bar{g}_{\mu\nu}$ have constant curvature, $\bar{R}_{\mu\nu}=\Lambda\bar{g}_{\mu\nu}$. 

Our next step is to derive the linearised equations around the constant curvature backgrounds. We decompose the metric into the background and a small fluctuation, $\gmn=\bar{g}_{\mu\nu}+\delta\gmn$, insert this decomposition into Einstein's equations and treat them perturbatively in $\delta \gmn$. At lowest order in the fluctuation, we recover the background equations which are identically satisfied by definition of $\bar{g}_{\mu\nu}$. At first order in $\delta \gmn$ we obtain the following equation,
\beqn\label{linmassless}
\bar{\mathcal{E}}_{\mu\nu}^{\phantom{\mu\nu}\rho\sigma}\delta g_{\rho\sigma}-\Lambda\Big(\delta\gmn-\frac{1}{2}\bar{g}_{\mu\nu}\bar{g}^{\rho\sigma}\delta g_{\rho\sigma}\Big)=0
\,,
\eeqn
where $\bar{\mathcal{E}}$ denotes the linearised Einstein operator in terms of the covariant derivative $\bar\nabla_\mu=\bar{g}_{\mu\rho}\bar\nabla^\rho$ compatible with the background metric,
\beqn\label{kinop}
	\bar{\mathcal{E}}_{\mu\nu}^{\phantom\mu\phantom\nu\rho\sigma} = \frac{1}{2}\Big(
	\delta^\rho_{~\mu}\delta^\sigma_{~\nu}\bar\nabla^2
	-\delta_{~\nu}^{\sigma}\bar\nabla_\mu\bar\nabla^\rho
	-\delta_{~\mu}^{\sigma}\bar\nabla_\nu\bar\nabla^\rho
	+\bar{g}_{\mu\nu} \bar\nabla^\sigma\bar\nabla^\rho
	+\bar{g}^{\rho\sigma}\bar\nabla_\mu\bar\nabla_\nu
	-\bar{g}_{\mu\nu}\bar{g}^{\rho\sigma}\bar\nabla^2
	\Big)
	\,.
\eeqn
It can be shown that equation~(\ref{linmassless}) describes a massless spin-2 field propagating on a maximally symmetric space-time. Out of the ten components in the symmetric field $\delta \gmn$ only two are physical while the remaining eight are non-dynamical, redundant degrees of freedom. This is a direct consequence of linearised coordinate invariance under which $\delta\gmn\mapsto\delta\gmn +\bar{\nabla}_\mu\xi_\nu+\bar{\nabla}_\nu\xi_\mu$. These gauge transformations leave~(\ref{linmassless}) invariant and are the direct analogue of the U(1) symmetry which removes the redundant degrees of freedom in the case of a massless spin-1 field $A_\mu$.

Note that the term proportional to the cosmological constant $\Lambda$ in~(\ref{linmassless}) is reminiscent of a mass term in the sense that it does not involve any derivatives. This interpretation  is wrong, however, because the term merely describes the correct coupling of the fluctuation to the background curvature. It vanishes in flat space, is gauge invariant and does not render the spin-2 field massive.

 % % % % % % % % % % % % % % % % % % % % % % % % % % % % % %
\subsection{Linear massive theory}
 % % % % % % % % % % % % % % % % % % % % % % % % % % % % % %
 
A mass term for the spin-2 field should not contain any derivatives. At the level of linear equations, the only possible terms are therefore $\delta\gmn$ and $\bar{g}_{\mu\nu}\bar{g}^{\rho\sigma}\delta g_{\rho\sigma}$. These two terms can however not be combined in an arbitrary way but their relative coefficient is fixed to be minus one.
Then the linearised equations for a massive spin-2 field propagating on a maximally symmetric spacetime read~\cite{Fierz:1939ix},
\beqn\label{linmass}
\bar{\mathcal{E}}_{\mu\nu}^{\phantom{\mu\nu}\rho\sigma}\delta g_{\rho\sigma}-\Lambda\Big(\delta\gmn-\frac{1}{2}\bar{g}_{\mu\nu}\bar{g}^{\rho\sigma}\delta g_{\rho\sigma}\Big)+\frac{m_\mathrm{FP}^2}{2}\Big(\delta\gmn-\bar{g}_{\mu\nu}\bar{g}^{\rho\sigma}\delta g_{\rho\sigma}\Big)=0
\,,
\eeqn
where the kinetic operator is the same as in~(\ref{kinop}). It can be shown that these equations propagate five degrees of freedom, corresponding to the five helicity states of the massive spin-2 field. If the relative coefficient between the two contributions to the mass term is changed, there is an extra dynamical scalar present in the theory which is a so-called ghost and gives rise to an instability. This pathology is associated with a wrong sign in front of the kinetic term for the scalar and therefore leads to a Hamiltonian which is unbounded from below. The correct linear mass term is tuned to produce a constraint that removes the ghost mode.

The massless linear spin-2 equations can be completed to the nonlinear Einstein equations of GR. This completion can partly be applied to~(\ref{linmass}) since the kinetic (and cosmological constant) terms are exactly the same as in~(\ref{linmassless}). The question is whether it is also possible to find a nonlinear completion for the mass term. 
It turns out that adding higher order non-derivative interactions for $\delta \gmn$ with generic coefficients to~(\ref{linmass}) does not result in a consistent nonlinear mass potential. In addition to the five degrees of freedom expected for a massive spin-2 field, a scalar mode enters the physical spectrum and gives rise to the so-called Boulware-Deser ghost instability~\cite{Boulware:1973my}. This ghost mode is the same that was eliminated from the linear theory by choosing the correct relative coefficient in the mass term. At the nonlinear level it generically returns and renders the theory inconsistent. In fact, Boulware and Deser claimed that no tuning of the nonlinear interaction parameters can give rise to a constraint that removes this mode. This conclusion turned out to be incorrect, however and there is a specific type of nonlinear mass potential that does avoid the ghost. Before introducing the ghost-free theory, we make a few general remarks on the structure of spin-2 interactions.

 % % % % % % % % % % % % % % % % % % % % % % % % % % % % % %
\subsection{General structure of the nonlinear theory}
 % % % % % % % % % % % % % % % % % % % % % % % % % % % % % % 

When trying to construct a mass term for the nonlinear tensor field $\gmn$, one immediately encounters the following obstruction: A nonlinear potential in the action must not have any lose indices, so we need an object to contract the indices of $\gmn$. This cannot be the metric itself since $g^{\mu\nu}$ with upper indices is the inverse metric and hence $g^{\mu\rho}g_{\rho\nu}=\delta^\mu_{~\nu}$, which cannot produce a nontrivial potential. Note that this problem does not arise for lower spin fields. For instance, one can straightforwardly build a mass term for the spin-1 vector $A_\mu$ using the inverse metric, $g^{\mu\nu}A_\mu A_\nu$. In the spin-2 case we are forced to introduce a new object to write down a mass term for $\gmn$. The simplest option is to invoke another tensor field which we shall call $\fmn$. The potential in the nonlinear action will then be a function of $f^{\mu\rho}g_{\rho\nu}$ (or, equivalently, its inverse $g^{\mu\rho}f_{\rho\nu}$) and we thus expect the ``massive gravity" action to have the following general structure,
\beqn\label{mggen}
S_\mathrm{mass}= m_\mathrm{P}^2\int\dd^4x~\sqrt{g}~(R-2\Lambda)-2m^4\int \dd^4 x~V(g,f)\,.
\eeqn
An example for a potential would be $V(g,f)=\sqrt{g}\,f^{\mu\rho}g_{\rho\mu}=\sqrt{g}\,\Tr(f^{-1}g)$, where we have included the scalar density $\sqrt{g}$ in order to give $V$ the correct transformation properties under GCT. However, this particular interaction term re-introduces the ghost instability and the same holds true for any generic form of $V$. The unique structure that avoids the ghost mode will be presented and discussed in the next section. 

The introduction of the second tensor field (also called ``reference metric") rises several questions. In the above action, $\fmn$ is non-dynamical, i.e.~it does not satisfy any equations of motion but is put into the theory by hand. It is therefore unclear how to determine the preferred form of $\fmn$, which may not seem satisfactory from a field theoretical point of view. For this reason, it may be preferable to also include a kinetic term for $\fmn$ into the action and thus work with a so-called bimetric theory of the form,
\beqn\label{bigen}
S_\mathrm{bi}= m_g^2\int\dd^4x~\sqrt{g}~(R(g)-2\Lambda)+m_f\int\dd^4x~\sqrt{f}~(R(f)-2\tilde\Lambda)-2m^4\int \dd^4 x~V(g,f)\,.
\eeqn
This is simply an extension of (\ref{mggen}) by a second Einstein-Hilbert term, where now $m_g$ and $m_f$ are the respective Planck masses for the two tensor fields. Bimetric theory faces the same problem as massive gravity with non-dynamical reference metric: For generic interaction potential $V$ an extra ghost mode makes the setup inconsistent.

%%%%%%%%%%%%%%%%%%%%%%%%%%%%%%%%%%%%%%%%%%%%%%%%%%%%%%%%%%%%%%%%
\section{The ghost-free theories}
%%%%%%%%%%%%%%%%%%%%%%%%%%%%%%%%%%%%%%%%%%%%%%%%%%%%%%%%%%%%%%%%

 % % % % % % % % % % % % % % % % % % % % % % % % % % % % % %
\subsection{Action and equations of motion}
 % % % % % % % % % % % % % % % % % % % % % % % % % % % % % % 
 
A unique structure of non-derivative spin-2 interactions gives rise to a constraint eliminating the unwanted ghost instability. It was proposed first for flat reference metric~\cite{deRham:2010ik, deRham:2010kj} and demonstrated to be ghost-free in this particular case~\cite{Hassan:2011hr}. The generalisation of the consistency proof to general and dynamical reference metrics was carried out shortly after~\cite{Hassan:2011tf, Hassan:2011zd} and established the existence of a bimetric theory that avoids the Boulware-Deser ghost.
Its potential reads,
\beqn\label{pot}
V(g,f)=\sqrt{g}~\sum_{n=1}^3 \beta_n e_n\big(\sqrt{g^{-1}f}\big)\,,
\eeqn
which includes three arbitrary interaction parameters $\beta_n$ and the elementary symmetric polynomials $e_n(S)$ which in terms of their matrix argument $S$ are defined as,
\beqn
e_1(S)&=&\Tr\,S\,,
\qquad
e_2(S)~=~\frac{1}{2}\Big(\big(\Tr\,S\big)^2-\Tr\big(S^2\big)\Big)\,,\nn\\
e_3(S)&=&\frac{1}{6}\Big(\big(\Tr\,S\big)^3-3\,\Tr\big(S^2\big)\big(\Tr\,S\big)+2\,\Tr\big(S^3\big)\Big)\,.
\eeqn
The argument $S$ that appears in the ghost free structure is the square-root of the matrix $g^{-1}f$, namely $S=\sqrt{g^{-1}f}$, defined through $S^2=g^{-1}f$. 

The potential~(\ref{pot}) can be inserted in the massive gravity action~(\ref{mggen}) or the bimetric action~(\ref{bigen}) and in both cases this results in a consistent theory. However, the physical content of the two setups is not identical: Massive gravity describes nonlinear self-interactions of a single massive spin-2 field whereas bimetric theory also contains a massless spin-2 field that mixes nonlinearly with the massive mode. We will analyse this in more detail in section~\ref{sec:mass}. From now on we shall focus on the fully dynamical bimetric theory which, as we will show in section~\ref{sec:gr}, unlike massive gravity possesses a smooth general relativity limit that is valid at all energy scales.

Varying the bimetric action~(\ref{bigen}) with respect to $\gmn$ and $\fmn$, results in two sets of equations of motion. In terms of the Einstein tensor, $\mathcal{G}_{\mu\nu}(g)=R_{\mu\nu}(g)-\frac{1}{2}\gmn R(g)$, which arises from the Einstein-Hilbert term, they read as,
\beqn
	\mathcal{G}_{\mu\nu}(g)+\Lambda\gmn+\frac{m^4}{m_g^{2}}V^g_{\mu\nu}(g,f)
	&=&0\,,\label{geom}
\\
	\mathcal{G}_{\mu\nu}(f)+\tilde\Lambda\fmn+\frac{m^4}{m_f^{2}}V^f_{\mu\nu}(g,f)
	&=&0\,.\label{feom}
\eeqn
The contributions $V^g_{\mu\nu}$ and $V^f_{\mu\nu}$ are derived from varying the interaction potential. Their explicit form is,
\beqn\label{potconbim}
	V^g_{\mu\nu}(g,f)&=&g_{\mu\rho}\sum_{n=1}^3(-1)^n\beta_n(Y_{(n)})^\rho_{~\nu}(S)\,,\\
	V^f_{\mu\nu}(g,f)&=&f_{\mu\rho}\sum_{n=1}^3(-1)^n\beta_{4-n}(Y_{(n)})^\rho_{~\nu}(S^{-1})\,,
\eeqn
where we have shortened the expressions by defining the following matrix-valued functions,
\beqn\label{yndef}
(Y_{(n)})^\rho_{~\nu}(S)\equiv\sum_{k=0}^n(-1)^k e_k(S)\,(S^{n-k})^\rho_{~\nu}\,.
\eeqn
In massive gravity with only one dynamical metric $\gmn$, the second set of equations (\ref{feom}) does not exist. 

We can now proceed and find solutions to the above equations of motion. Note that all our considerations here are in vacuum; matter couplings will be discussed in section~\ref{sec:gr}. We therefore focus on vacuum solutions which constitute the analogue of metrics satisfying $R_{\mu\nu}=\Lambda\gmn$ in GR. Due to the interaction potential the solution spectrum of bimetric theory is richer than that of GR, but for now we restrict ourselves to the simplest class of maximally symmetric backgrounds satisfying (\ref{geom}) and (\ref{feom}).

% % % % % % % % % % % % % % % % % % % % % % % % % % % % % %
\subsection{Proportional backgrounds}\label{sec:pbg}
% % % % % % % % % % % % % % % % % % % % % % % % % % % % % %
Following the analysis of~\cite{Hassan:2012wr}, we  make the following ansatz for the two tensor fields,
\beqn
\fmn=c^2\gmn\,,
\eeqn 
where $c$ is an arbitrary (real or imaginary) constant.\footnote{In principle we could make the same ansatz with a scalar function $c(x)$ but the equations of motion will immediately imply that $c(x)$ is constant.} Since on this ansatz we have $g^{\mu\rho}f_{\rho\nu}=c^2 \delta^\mu_{~\nu}$, the bimetric equations (\ref{geom}) and (\ref{feom}) reduce to,
\beqn
	\mathcal{G}_{\mu\nu}(g)+\Lambda_g\gmn
	&=&0\,,\label{geomprop}
\\
	\mathcal{G}_{\mu\nu}(c^2 g)+\Lambda_f\gmn
	&=&0\,,\label{feomprop}
\eeqn
where the effective cosmological constant contributions are,
\beqn
\Lambda_g&=&\Lambda+\frac{m^4}{m_g^{2}}\big(3c\beta_1+3c^2\beta_2+c^3\beta_3\big)\,,\label{ccg}\\
\Lambda_f&=&\tilde\Lambda+\frac{m^4}{c^2m_f^{2}}\big(c\beta_1+3c^2\beta_2+3c^3\beta_3\big)\label{ccf}\,.
\eeqn
Hence, on the proportional ansatz the bimetric equations simply become two copies of Einstein's equation.
Since the Einstein tensor is scale invariant, $\mathcal{G}_{\mu\nu}(c^2g)=\mathcal{G}_{\mu\nu}(g)$, by comparing the left-hand sides of (\ref{geomprop}) and (\ref{feomprop}) we obtain the constraint,
\beqn\label{ccc}
\Lambda_g=\Lambda_f\,.
\eeqn
This is a quartic polynomial equation for the proportionality constant $c$ and solving it fully specifies the solution.\footnote{In an exceptional case, for particular values of $\Lambda$, $\tilde\Lambda$ and $\beta_n$, eq.~(\ref{ccc}) is identically satisfied and does not determine~$c$. This particular bimetric potential has been discussed in the context of nonlinear partial masslessness~\cite{Hassan:2012gz}, see also~\cite{deRham:2013wv}.} 

We conclude that the proportional background solutions in bimetric theory are a direct analogue of vacuum solutions in GR, since they are characterised by constant curvatures for both metrics, $R_{\mu\nu}(g)=\Lambda_g\gmn$ and $R_{\mu\nu}(f)=\Lambda_fc^{-2}\fmn$. It turns out that these bimetric solutions are the only backgrounds around which the perturbation equations can be diagonalised in terms of spin-2 mass eigenstates. 

Before discussing the linearised equations and the mass spectrum in more detail in the next subsection, let us point out a crucial difference between bimetric theory and GR. The vacuum energy $\Lambda_\mathrm{vac}$ in GR arises from quantum loops in the matter sector and combines with the bare cosmological constant $\Lambda_0$ into the fully renormalised vacuum energy contribution $\Lambda=\Lambda_\mathrm{vac}+\Lambda_0$ in Einstein's equations~(\ref{EE}). The combined quantity determines the background curvature and is measured to have a very tiny, positive value. It is often argued that this observation is in conflict with theoretical expectations since quantum field theory computations suggest a value for $\Lambda_\mathrm{vac}$ that is much larger. This means that also the bare cosmological constant $\Lambda_0$ has to be large and cancel $\Lambda_\mathrm{vac}$ to many decimal places in order to yield the correct value for the measured background curvature $\Lambda$. Many theorists are unsatisfied with this picture because they regard such a cancellation as ``fine-tuned" and ``unnatural".

In bimetric theory the situation is slightly different since, in addition to $\Lambda=\Lambda_\mathrm{vac}+\Lambda_0$, the observed value $\Lambda_g$ also receives contributions from the interaction potential, c.f.~equation (\ref{ccg}). This means that if there existed a mechanism ensuring that $\Lambda=0$ (such as supersymmetry, if realised at all energies, for instance), then the only contribution would be $\Lambda_\mathrm{int}=\frac{m^4}{m_g^{2}}\big(3c\beta_1+3c^2\beta_2+c^3\beta_3\big)$, coming from the spin-2 interactions. The advantage of this scenario is that the mass parameter $m$ is protected against receiving large quantum corrections and $\Lambda_\mathrm{int}$ does therefore not face the same problem as the bare cosmological constant $\Lambda_0$. This type of protection is at work due to an enhancement of symmetry (to wit, full general coordinate invariance for both metrics) in the massless theory with $m=0$.

In this sense, bimetric theory can solve part of the problem associated to vacuum energy in GR. It cannot explain why the bare cosmological constant and the contribution from matter loops cancel each other exactly. However, if we assume that this is the case, it provides an additional type of vacuum energy whose value is ``technically natural" in the sense that it does not receive large corrections from renormalisation. The existence of this feature is not restricted to the proportional backgrounds but can also be verified for more realistic homogeneous and isotropic solutions. We shall come back to this in section~\ref{sec:appl}.

% % % % % % % % % % % % % % % % % % % % % % % % % % % % % %
\subsection{Mass spectrum}\label{sec:mass}
% % % % % % % % % % % % % % % % % % % % % % % % % % % % % %

The presence of the square-root matrix which is crucial for avoiding the ghost complicates most calculations in massive gravity and bimetric theory. In particular, deriving linear perturbation equations is much more involved than in GR. For general backgrounds, these equations have recently been presented and analysed in~\cite{Bernard:2015uic}, but their structure is not very illustrative. Fortunately, around the proportional background solutions the situation drastically simplifies. Plugging the decompositions $\gmn=\bar{g}_{\mu\nu}+\delta\gmn$ and $\fmn=c^2\bar{g}_{\mu\nu}+\delta\fmn$ into (\ref{geom}) and (\ref{feom}), it is straightforward to obtain the linearised equations for the fluctuations $\delta\gmn$ and $\delta\fmn$.  The resulting expressions mix $\delta\gmn$ and $\delta\fmn$ which are therefore not mass eigenstates. However, we can build two linear combinations of the equations in order to decouple them. The result reads,
\beqn
{\bar{\mathcal{E}}_{\mu\nu}}^{~~\rho\sigma}\delta G_{\rho\sigma}-{\Lambda}_g\Big(\delta G_{\mu\nu}-\frac{1}{2}\bar{g}_{\mu\nu}\delta G\Big) &=&0\,, \label{dGeq}\\
{\bar{\mathcal{E}}_{\mu\nu}}^{~~\rho\sigma}\delta M_{\rho\sigma}-{\Lambda}_g\Big(\delta M_{\mu\nu}-\frac{1}{2}\bar{g}_{\mu\nu}\delta M\Big)
+\frac{{m}^2_\mathrm{FP}}{2}\Big(\delta M_{\mu\nu}-\bar{g}_{\mu\nu}\delta M\Big)&=&0\,,\label{dMeq}
\eeqn
where the kinetic operator ${\bar{\mathcal{E}}_{\mu\nu}}^{~~\rho\sigma}$ is the same as in~(\ref{kinop}).
Comparing these equations to (\ref{linmassless}) and (\ref{linmass}), we see that the first set describes a massless and the second a massive spin-2 field on maximally symmetric backgrounds. 
Expressed in terms of the original fluctuations, the mass eigenstates which diagonalise the equations are given by,
\beqn\label{dGanddM}
\delta G_{\mu\nu} =\delta g_{\mu\nu}+\alpha^{2}\delta f_{\mu\nu}\,,\qquad
\delta M_{\mu\nu} = \frac{1}{2c}\big(\delta f_{\mu\nu}-c^2 \delta g_{\mu\nu}\big)\,,
\eeqn 
where we have used the abbreviation $\alpha\equiv m_f/m_g$. The Fierz-Pauli mass is a function of the parameters of the theory,
\beqn\label{FPmass}
{m}^2_\mathrm{FP}=
\frac{m^4}{m_g^{2}} \big(1+\alpha^{-2}c^{-2}\big)\left(c\beta_1 +2c^2\beta_2+c^3\beta_3\right)\,.
\eeqn
In all these expressions, $c$ is to be regarded as a function of bimetric parameters as well, determined by the background condition (\ref{ccc}).

% % % % % % % % % % % % % % % % % % % % % % % % % % % % % %
\subsection{General relativity limit}\label{sec:gr}\label{sec:GR}
% % % % % % % % % % % % % % % % % % % % % % % % % % % % % % 

So far we have only considered bimetric theory in vacuum, ignoring interactions with other matter fields. A priori, since we are dealing with two tensor fields, there are many possibilities for coupling the gravitational sector to matter. However, it turns out that most of them  re-introduce the ghost instability. The only known consistent couplings are of the same form as in GR, i.e.~two copies of the Lagrangian in equation~(\ref{matter}),
\beqn\label{tmc}
S_\mathrm{m}=\int\dd^4x~\mathcal{L}_\mathrm{m}^g(g, \Phi^g_i)+\int\dd^4x~\mathcal{L}_\mathrm{m}^f(f, \Phi^f_i)\,.
\eeqn
Here it is crucial that the matter fields in the two couplings must be different, $\Phi^g_i\neq\Phi^f_j$. Coupling $\gmn$ and $\fmn$ to the same matter fields will again excite the Boulware-Deser ghost~\cite{Yamashita:2014fga, deRham:2014naa}. We can therefore imagine that the Standard Model directly interacts only with one ``physical" metric, say $\gmn$, and sees the second tensor $\fmn$ only indirectly through the dynamics of $\gmn$. In the following, for simplicity and to keep the number of new ingredients minimal, we shall assume that the matter sector of $\fmn$ is empty, $\mathcal{L}_\mathrm{m}^f=0$. Then the bimetric equations with matter source for $\gmn$ read,
\beqn
	\mathcal{G}_{\mu\nu}(g)+\Lambda\gmn+\frac{m^4}{m_g^{2}}V^g_{\mu\nu}(g,f)
	&=&\frac{1}{m_g^2}T_{\mu\nu}\,,\label{mgeom}
\\
	\mathcal{G}_{\mu\nu}(f)+\tilde\Lambda\fmn+\frac{m^4}{m_f^{2}}V^f_{\mu\nu}(g,f)
	&=&0\,,\label{mfeom}
\eeqn
where the stress-energy tensor is again defined as $T_{\mu\nu}=-\frac{1}{\sqrt{g}}\frac{\delta \mathcal{L}_\mathrm{m}^g}{\delta g^{\mu\nu}}$.

We will now discuss the parameter limit for which these equations approach GR. From a phenomenological point of view, it is interesting to have a parameter that can give us a feeling for how much the predictions of bimetric theory differ from those of the well-tested standard framework. One may expect this parameter to be the mass of the massive spin-2 mode, which is proportional to $m$ but, as known for a long time already, the $m\rightarrow 0$ limit for massive spin-2 fields is not continuous~\cite{vanDam:1970vg, Zakharov:1970cc}. In other words, by taking the mass to zero, we do not recover the massless theory and therefore, for bimetric theory or massive gravity, this limit does not result in GR. In fact, massive gravity possesses no GR limit valid at all energy scales\footnote{Nonlinear massive gravity can exhibit the Vainshtein mechanism~\cite{Vainshtein:1972sx} which allows for a smooth transition to GR in certain energy regimes.} and in bimetric theory we have to perform some nontrivial manipulations in order to identify the correct parameter scaling. The starting point is the following identity satisfied by the potential contributions in (\ref{mgeom}) and (\ref{mfeom})~\cite{Baccetti:2012bk, Hassan:2014vja},
\beqn\label{Vid}
\sqrt{g}~g^{\mu\rho}V^g_{\rho\nu}+\sqrt{f}~f^{\mu\rho}V^f_{\rho\nu}-\sqrt{g}~V\delta^\mu_{~\nu}=0\,,
\eeqn
where $V=V(g,f)$ is the potential~(\ref{pot}) as it appears in the action. Using this identity we can combine the equations of motion (\ref{mgeom}) and (\ref{mfeom}) to obtain the following equation,
\beqn\label{combeq}
g^{\mu\rho}\mathcal{G}_{\rho\nu}(g)+\alpha^2\det\big(\sqrt{g^{-1}f}\,\big)\,f^{\mu\rho}\mathcal{G}_{\rho\nu}(f)+\frac{m^4}{m_g^2} V \delta^\mu_{~\nu}=\frac{1}{m_g^2}g^{\mu\rho}T^g_{\rho\nu}\,,
\eeqn
where $\alpha=m_f/m_g$, as before. Now consider the limit $\alpha\rightarrow0$ for which this equation reduces to,
\beqn
\mathcal{G}_{\mu\nu}(g)+\frac{m^4}{m_g^2} V \gmn=\frac{1}{m_g^2}T^g_{\mu\nu}\,.
\eeqn
Taking the divergence of both sides with the covariant derivative $\nabla$ compatible with $\gmn$ and using the Bianchi identity, $\nabla^\mu\mathcal{G}_{\mu\nu}(g)=0$, we find the on-shell constraint $\partial_\nu V=0$ and hence $V=$\,const. This implies that, in the limit $\alpha\rightarrow0$, the physical metric $\gmn$ satisfies an Einstein equation, $\mathcal{G}_{\mu\nu}(g)+\Lambda\gmn=0$, with cosmological constant $\Lambda=\frac{m^4}{m_g^2} V$. We conclude that the GR limit of bimetric theory corresponds to taking,
\beqn 
\alpha\longrightarrow 0\,, \qquad
m_g=\mathrm{const.}\,, \qquad 
T^f_{\mu\nu}=-\frac{1}{\sqrt{f}}\frac{\delta \mathcal{L}_\mathrm{m}^f}{\delta f^{\mu\nu}}=0\,.
\eeqn
A complementary way of deriving the GR limit at the level of perturbations is to look at equation~(\ref{dGanddM}). For $\alpha\rightarrow 0$ we see that $\delta G_{\mu\nu}\rightarrow \delta \gmn$ and hence the fluctuation that couples to matter becomes massless, as it should in GR. In other words, taking $\alpha$ to zero decouples the massive spin-2 mode from the matter sector instead of sending its mass to zero. For this reason, the so-defined limit does not have the discontinuity problems of the usual $m\rightarrow 0$ approach and it is valid also for large values of the spin-2 mass. 

Interestingly, a small value for $\alpha$ automatically corresponds to a small inverse mass $m_g^{-1}$ which sets the scale for the gravitational interactions of matter fields. The GR limit is thus compatible with the fact that the observed value for the physical Planck mass is large, i.e.~gravity is weak. For instance, say we take $m_g$ to be on the order of the observed Planck mass and $m_f$ to be on the order of the weak scale; this results in a tiny value for $\alpha$ and does not create any new hierarchy between energy scales.

Before concluding this section let us make one last remark.
In the $\fmn$ equations~(\ref{mfeom}), taking $\alpha$ to zero eliminates the kinetic term. Note that we have the freedom to impose a scaling of $\tilde\Lambda$ such that it survives in the limit of vanishing $\alpha$. The equations become algebraical and the generic solution for $\fmn$ is the proportional one, $\fmn=c^2\gmn$, in which $c$ is determined by the equation $\Lambda_f=0$, where $\Lambda_f$ was defined in (\ref{ccf}). One may be worried about strong-coupling of the $\fmn$ fluctuations in the limit of small but finite $\alpha$. However, as discussed in~\cite{Akrami:2015qga}, this problem does not arise for a large range of phenomenologically viable values for $\alpha$.

%%%%%%%%%%%%%%%%%%%%%%%%%%%%%%%%%%%%%%%%%%%%%%%%%%%%%%%%%%%%%%%%
\section{Applications to cosmology}\label{sec:appl}
%%%%%%%%%%%%%%%%%%%%%%%%%%%%%%%%%%%%%%%%%%%%%%%%%%%%%%%%%%%%%%%%

Generically, massive gravity and bimetric theory are expected to make predictions for gravitational phenomena that differ from those of GR. The most interesting question is whether the novel theories are able to address the dark energy, dark matter or the cosmological constant problems without violating tests of GR at smaller distance scales. 

A lot of effort has been spent on the study of cosmological solutions in bimetric theory.\footnote{As before we will focus on the fully dynamical case because massive gravity with fixed reference metric cannot give rise to viable cosmologies; see~\cite{D'Amico:2011jj}, for instance.}
Before we summarise their features, let us very briefly review some of the main ingredients of standard cosmology in GR.
Evaluating Einstein's equations (\ref{EE}) on a homogeneous and isotropic ansatz for the metric $\gmn$ and the stress-energy tensor $T_{\mu\nu}$, one obtains the Friedmann equation,\footnote{For the sake of notational simplicity, we have assumed the spatial curvature of the metric to vanish, $k=0$. This assumption has no impact on the discussions in the following.}
\beqn\label{FE}
\left(\frac{\dot{a}}{a}\right)^2= \frac{\Lambda}{3}+\frac{\rho(t)}{3m_\mathrm{P}^2}\,.
\eeqn
Here, $a(t)$ is the scale factor describing the expansion of space, the dot denotes a time derivative and $\rho(t) = -T^0_{~0}$ is the energy density of the matter sector. The latter is found to be dominated by a dark matter component consisting of unknown particles that are not included in the Standard Model.  
Moreover, another component of Einstein's equations reveals that the presence of a small cosmological constant $\Lambda$ is necessary to achieve $\ddot{a}>0$ and thus describe the observed accelerated expansion. As already mentioned in the introduction, this gives rise to two major problems: (i) One needs to explain what happens to the large value for $\Lambda$ that one would expect from quantum field theory computations. This puzzle is usually referred to as the cosmological constant problem. (ii) Assuming that the large vacuum energy contribution is somehow cancelled or rendered irrelevant, one still needs to address the dark energy problem and explain the origin of the small observed $\Lambda$.

\paragraph{The cosmological constant problem.}
One of the original hopes was that a massive graviton could serve to ``screen out" a large vacuum energy contribution coming from the matter sector. The massless spin-2 field in GR gives rise to a Newtonian potential that decays as $1/r$ with distance~$r$. Introducing a Fierz-Pauli mass $m_\mathrm{FP}$ for the graviton, on the other hand, would lead to a Yukawa potential of the shape $e^{-m_\mathrm{FP}r}/r$. At small distances $r\ll m_\mathrm{FP}$, the Yukawa potential behaves like the Newtonian one but at large distances the exponential decay sets in. This weakens the gravitational force at large distances and, as a direct consequence, vacuum energy has a smaller impact on the expansion rate of the universe. Unfortunately, it turns out that this promising feature of linear massive gravity is not realised in the theories describing nonlinear interactions. It thus seems that bimetric theory and massive gravity are not able to address the cosmological constant problem (i) and in this respect have the same status as GR.

\paragraph{The dark energy problem.}
First studies of cosmology in bimetric theory were performed in \cite{Volkov:2011an, vonStrauss:2011mq, Comelli:2011zm} where the equations of motion including matter, (\ref{mgeom}) and (\ref{mfeom}), were solved for homogeneous and isotropic ans\"atze for both metrics. The generic outcome is a modified Friedmann equation for the scale factor $a(t)$ of the metric $\gmn$,
\beqn\label{mFE}
\left(\frac{\dot{a}}{a}\right)^2=\frac{\Lambda}{3}+ F\big[\rho(t)\big]\,.
\eeqn
Here $F\big[\rho(t)\big]$ is a function of the energy density $\rho(t)$, which in the GR case (\ref{FE}) was simply linear. For generic interaction parameters in bimetric theory, the function will now be nonlinear and hence the cosmological background evolution will differ significantly from that of GR. This means that a large region in the bimetric parameter space can immediately be ruled out since it does not describe the observed evolution correctly. 

The crucial property of many viable background solutions in bimetric theory is that acceleration is possible even in the absence of vacuum energy, i.e.~for $\Lambda=0$. We already encountered this feature in the case of proportional backgrounds in section~\ref{sec:pbg} and are now ready to discuss it in the context of the modified Friedmann equation (\ref{mFE}).
For instance, if one takes $\beta_1=\beta_3=0$ in the bimetric interaction potential, then $F\big[\rho(t)\big]$ reduces to a constant and hence acts as vacuum energy in (\ref{mFE})~\cite{vonStrauss:2011mq}. Although the bimetric equations evaluated on the homogeneous and isotropic ansatz produce the ordinary Friedmann equation in this case, the full theory is not equivalent to GR. Linear perturbations and nonlinear interactions differ from the standard scenario and can give rise to observable signatures. For more general parameters, $F\big[\rho(t)\big]$ may have non-constant contributions as well but, if their effects are sufficiently suppressed, then (\ref{mFE}) is able to describe the background evolution in cosmology just as well as the standard Friedmann equation (\ref{FE}). A full scan of the bimetric parameter space with $\Lambda=0$ has been performed in~\cite{Akrami:2012vf} and confirms the existence of a large set of self-accelerating cosmological solutions.\footnote{In fact, de Sitter solutions always constitute a late-time attractor in the modified evolution equations, provided that the energy density for matter decays with increasing scale factor as usual~\cite{vonStrauss:2011mq}.}

The scale of cosmic acceleration in bimetric models with $\Lambda=0$ is thus set by the spin-2 mass scale $m$ which is protected against receiving large quantum corrections. As stated earlier, the protection mechanism is a consequence of the fact that taking $m$ to zero restores the general coordinate invariance of the massless theory. On the other hand, no symmetry is recovered in the $\Lambda\rightarrow 0$ limit and vacuum energy is therefore not a protected quantity.
In this sense, when compared to GR, bimetric theory has the advantage that its dark energy scale can be small and still be ``technically natural". However, as outlined above, the theory cannot explain what happens to the large amount of vacuum energy that is expected to arise from the matter sector.

Linear perturbations around most branches of cosmological solutions give rise to various pathologies, see for instance~\cite{Konnig:2015lfa}. One branch is well-behaved but some of its perturbations become strongly coupled at early times, leading to a breakdown of linear perturbation theory~\cite{Comelli:2012db, Konnig:2014xva}. In spite of avoiding the consistency problems, the models lose their predictivity and studying their phenomenology requires the development of nonlinear methods.\footnote{For large values of the spin-2 mass scale $m$, this problem does not occur~\cite{DeFelice:2014nja}. However, in this case the self-acceleration scale can no longer be small in a natural way.}
Fortunately, this obstacle can be avoided in the GR limit: For small values of $\alpha=m_f/m_g$, the time at which linear perturbation theory loses its validity is pushed so far into the past that the issue becomes irrelevant for observational phenomena~\cite{Akrami:2015qga}. Interestingly, even for $\Lambda=0$ the function $F[\rho(t)]$ in (\ref{mFE}) contains a constant contribution which is independent of $\alpha$. The self-acceleration mechanism therefore survives in the GR limit and bimetric theory with small $\alpha$ predicts a viable cosmology whose dark energy scale is protected against quantum corrections.

\paragraph{The dark matter problem.}
The literature does not contain much work on trying to explain the nature of dark matter with the help of bimetric theory and what we are going to present here is rather speculative. An idea recently pushed forward in~\cite{Aoki:2014cla} involved coupling different types of matter to the two metrics as in equation~(\ref{tmc}).\footnote{A similar model has been studied in~\cite{Blanchet:2015bia}, but it has a ghost and is therefore inconsistent in its present form.} This setup seems indeed successful in reproducing some of the observations related to dark matter. 

We now suggest a more minimal approach without introducing any extra fields or inconsistent couplings. Once more we shall assume that only $\gmn$ interacts with matter sources as in~(\ref{mgeom}).
As we already mentioned in section~\ref{sec:GR}, in the GR limit with $\alpha\rightarrow 0$, the massive spin-2 mode $\delta M_{\mu\nu}$ decouples from the matter sector. At the linear level this is obvious because the fluctuation $\delta \gmn$ of the physical metric approaches the massless mode $\delta G_{\mu\nu}$ in this limit. For small but finite $\alpha$, the interactions of matter with $\delta M_{\mu\nu}$ are suppressed by powers of $\alpha$ since from (\ref{dGanddM}) we have that, 
\beqn
\delta\gmn=\frac{1}{1+\alpha^2c^2}(\delta G_{\mu\nu}-2\alpha^2c\delta M_{\mu\nu})\,.
\eeqn
In the nonlinear matter coupling, each vertex with $n$ powers of $\delta \gmn$ therefore leads to interactions of $\delta M_{\mu\nu}$ with matter fields (and $\delta G_{\mu\nu}$) which are suppressed by coefficients ranging from $\alpha^2$ to $\alpha^{2n}$. Bimetric theory with small $\alpha$ can thus be interpreted as a model for gravity whose dynamics and interactions with matter fields are very close to those of GR. The difference lies in the presence of an additional massive spin-2 field that mixes with the massless graviton at the nonlinear level but interacts very weakly with matter. The mass of this extra field is a free parameter of the theory and, even if it is taken to be large, it does not make the gravitational interactions of ordinary matter differ significantly from GR.
In principle it is possible that dark matter is composed out of this massive spin-2 field which gravitates but couples only very weakly to matter. It would therefore be very interesting to study the phenomenology of this scenario and obtain predictions from the spin-2 model for dark matter related observables.

%%%%%%%%%%%%%%%%%%%%%%%%%%%%%%%%%%%%%%%%%%%%%%%%%%%%%%%%%%%%%%%%
\section{Summary and outlook}
%%%%%%%%%%%%%%%%%%%%%%%%%%%%%%%%%%%%%%%%%%%%%%%%%%%%%%%%%%%%%%%%

Ghost-free bimetric theory is one of the few consistent modifications of general relativity. The particular structure of its interaction potential with only three free parameters avoids the notorious Boulware-Deser ghost mode which generically plagues all types of nonlinear interactions for massive spin-2 fields. The theory propagates a massive and a massless spin-2 mode around its maximally symmetric backgrounds. These modes mix heavily at the nonlinear level and the physical metric that couples to matter is neither massless nor massive. However, we have seen that there exists a limit in the parameter space of bimetric theory which decouples the massive spin-2 field from the matter sector and GR is smoothly recovered in the exact limit. We discussed how this limit can be invoked to produce a viable cosmology with a self-acceleration scale that is protected against receiving large quantum corrections. At the same time, since its interactions with matter are very weak, the massive spin-2 mode could potentially be a candidate for dark matter. 

There are still a lot of open questions concerning the interacting theories for massive spin-2 fields. On the phenomenological side, we already mentioned the possible application to the dark matter problem where little work has been done so far. More generally, it is of great importance to identify observable signatures for the presence of the massive mode in order to distinguish bimetric theory from GR.
One possibility could be to study bimetric theory in the context of inflation and investigate the impact of massive spin-2 fluctuations on the CMB power spectrum.

On the theoretical side, it is very likely that bimetric theory in its present form does not yield the complete description of massive spin-2 fields. 
At the quantum level one would expect that a new type of symmetry breaking mechanism (analogous to the Higgs mechanism for spin-1 fields) will be required for consistency. Such a mechanism has not been found yet but it will most likely require introducing new degrees of freedom.
Several other possibilities of further extending bimetric theory have already been excluded~\cite{deRham:2015cha}: For instance, it seems impossible to include new kinetic couplings for the metrics without reintroducing the ghost. Furthermore, in models with more than two spin-2 fields, the interactions can come only as copies of the ghost-free bimetric potential and vertices with more than two different fields are not allowed. Potential extensions of bimetric theory should therefore contain other fields, maybe even exotic ones with spin higher than two.

There exist still some unresolved issues within pure bimetric theory. For instance, a special parameter choice in the interaction potential results in a particularly interesting structure, possibly giving rise to an additional symmetry~\cite{Hassan:2015tba}. This feature seems to be related to the phenomenon of partial masslessness (referring to a spin-2 field with four propagating degrees of freedom on a de Sitter spacetime~\cite{Deser:2001pe}) and may eventually be relevant for the cosmological constant problem.

In summary, bimetric theory is a promising extension of standard gravity with interesting new features and therefore remains an exciting field of study.

\bigskip
{\bf Acknowledgements}: The author would like to thank the organisers of the ``Corfu Summer Institute 2015" for the invitation and Mikael von Strauss for helpful comments on the draft.
This work is supported by ERC grant no.~615203 under the FP7 and the Swiss National Science Foundation through the NCCR SwissMAP.

%%%%%%%%%%%%%%%%%%%%%%%%%%%%%%%%%%%%%%%%%%%%%%%%%%%%%%%%%%%%%%%%
%%%%%%%%%%%%%%%%%%%%%%%%%%%%%%%%%%%%%%%%%%%%%%%%%%%%%%%%%%%%%%%%
%%%%%%%%%%%%%%%%%%%%%%%%%%%%%%%%%%%%%%%%%%%%%%%%%%%%%%%%%%%%%%%%

\end{document}